\documentclass[prd,superscriptaddress,twocolumn,amsfonts,amssymb,amsmath,showpacs]{revtex4-2}
\usepackage{bm}
\usepackage{amsfonts}
\usepackage{latexsym}
\usepackage[latin1]{inputenc}
\usepackage{graphicx}
\usepackage{amsmath}
\usepackage{palatino}
\usepackage{mathpazo}
\usepackage[british]{babel}
\usepackage{hhline}
\usepackage{multirow}
\usepackage{textcomp}
\usepackage{float}
\usepackage{booktabs}
\usepackage{dcolumn}
\usepackage{hhline}
\usepackage{multirow}
\usepackage{ragged2e}
\usepackage{hyperref}
\hypersetup{colorlinks,citecolor=blue}
\hypersetup{colorlinks=true,linkcolor=red,filecolor=magenta,    urlcolor=cyan}
\usepackage{amsmath}
\usepackage{xcolor}
\usepackage{orcidlink}
\usepackage{epsfig}
\usepackage{caption}
\usepackage{subcaption}
\usepackage{commath}
\def\jnl@style{\it}
\def\aaref@jnl#1{{\jnl@style#1}}

\def\aaref@jnl#1{{\jnl@style#1}}

\def\aj{\aaref@jnl{AJ}}                   % Astronomical Journal
\def\apj{\aaref@jnl{ApJ}}                 % Astrophysical Journal
\def\apjl{\aaref@jnl{ApJ}}                % Astrophysical Journal, Letters
\def\apjs{\aaref@jnl{ApJS}}               % Astrophysical Journal, Supplement
\def\apss{\aaref@jnl{Ap\&SS}}             % Astrophysics and Space Science
\def\aap{\aaref@jnl{A\&A}}                % Astronomy and Astrophysics
\def\aapr{\aaref@jnl{A\&A~Rev.}}          % Astronomy and Astrophysics Reviews
\def\aaps{\aaref@jnl{A\&AS}}              % Astronomy and Astrophysics, Supplement
\def\mnras{\aaref@jnl{Mon.~Not.~Roy.~Astron.~Soc.}}             % Monthly Notices of the RAS
\def\prd{\aaref@jnl{Phys.~Rev.~D}}        % Physical Review D
\def\prc{\aaref@jnl{Phys.~Rev.~C}}  % Physical Review C
\def\prl{\aaref@jnl{Phys.~Rev.~Lett.}}    % Physical Review Letters
\def\qjras{\aaref@jnl{QJRAS}}             % Quarterly Journal of the RAS
\def\skytel{\aaref@jnl{S\&T}}             % Sky and Telescope
\def\ssr{\aaref@jnl{Space~Sci.~Rev.}}     % Space Science Reviews
\def\zap{\aaref@jnl{ZAp}}                 % Zeitschrift fuer Astrophysik
\def\nat{\aaref@jnl{Nature}}              % Nature
\def\aplett{\aaref@jnl{Astrophys.~Lett.}} % Astrophysics Letters
\def\apspr{\aaref@jnl{Astrophys.~Space~Phys.~Res.}} % Astrophysics Space Physics Research
\def\physrep{\aaref@jnl{Phys.~Rep.}}      % Physics Reports
\def\physscr{\aaref@jnl{Phys.~Scr}}       % Physica Scripta
\def\commat{\aaref@jnl{Comm.~Math.~Phys.}}              % Communications in Mathematical Physics
\def\science{\aaref@jnl{Science}}               % Science
\def\cqg{\aaref@jnl{Classical Quant.~Grav.}}            % Classical and Quantum Gravity
\def\jpcs{\aaref@jnl{JPCS}}                                     % Journal of Physics Conference Series
\def\ijmpd{\aaref@jnl{Int.~J.~Mod.~Phys.~D}}                    % International Journal of Modern Physics D
\def\grg{\aaref@jnl{Gen.~Relat.~Gravit.}}               % General Relativity and Gravitation
\def\rpp{\aaref@jnl{Rep.~Prog.~Phys.}}          % Reports on Progress in Physics
\def\npa{\aaref@jnl{Nucl.~Phys.~A}}        % Nuclear Physics A
\def\lrr{\aaref@jnl{Living Rev.~Rel.}}                   % Living reviews in relativity
\def\jcap{\aaref@jnl{J.~Cosmology Astropart.~Phys.}}    % Journal of cosmology and astroparticle physics
\def\rmp{\aaref@jnl{Rev.~Mod.~Phys.}}   %Reviews of modern physics
\def\epjc{\aaref@jnl{Eur.~Phys.~J.~C}} 
\def\plb{\aaref@jnl{~Phy.~Lett.~B}} 
\def\mpla{\aaref@jnl{Mod.~Phy.~Lett.~A}} 
\def\arxiv{\aaref@jnl{arxiv.org}}

\allowdisplaybreaks[1]
\begin{document}
%\color{red}
\color{black}       %% For one column
%
%\title{\bf Constraining $f(T,\mathcal{T})$ gravity from dynamical system analysis}
\title{\bf Cosmological models in $f (T, \mathcal{T})$ gravity and the dynamical system analysis}

%\end{document}

\author{L.K. Duchaniya \orcidlink{0000-0001-6457-2225}}
\email{duchaniya98@gmail.com}
\affiliation{Department of Mathematics,
Birla Institute of Technology and Science-Pilani, Hyderabad Campus,
Hyderabad-500078, India.}

\author{Santosh V Lohakare \orcidlink{0000-0001-5934-3428}}
\email{lohakaresv@gmail.com}
\affiliation{Department of Mathematics,
Birla Institute of Technology and Science-Pilani, Hyderabad Campus,
Hyderabad-500078, India.}

 \author{B. Mishra\orcidlink{0000-0001-5527-3565}}
 \email{bivu@hyderabad.bits-pilani.ac.in}
 \affiliation{Department of Mathematics, Birla Institute of Technology and Science-Pilani, Hyderabad Campus, Hyderabad-500078, India.}

\begin{abstract}
{\bf{Abstract}:} The study addresses matter-coupled modified gravity, particularly $f (T, \mathcal{T})$ gravity, unveiling distinct formalism. The research further discusses stability analysis, and the dynamical system approach, exploring the dynamics of critical points to understand these models' viability better. The dynamical system analysis of the cosmological models in $f(T, \mathcal{T})$ gravity, where $T$ and $\mathcal{T}$ respectively represent the torsion scalar and trace of the energy-momentum tensor has been investigated. It demonstrates how first-order autonomous systems can be treated as cosmological equations and analyzed using standard dynamical system theory techniques. Two forms of the function $f(T,\mathcal{T})$ are considered (i) one with the product of trace and higher order torsion scalar and the other (ii) linear combination of linear trace and squared torsion. By employing this methodology, the research aims to uncover the actual behavior of the Universe. The findings emphasize the graphical representation of these insights, enriching our understanding of cosmological scenarios.  
\end{abstract}

\maketitle
%\textbf{PACS number}: \\

\section{Introduction} \label{SEC-I}
 The presence of mysterious forms of energy in the Universe is claimed to be the reason for the accelerated expansion of the Universe \cite{Riess:1998cb, Perlmutter:1998np}. This mysterious energy has been termed dark energy (DE) and has been investigated from different perspectives, such as the cosmic microwave background (CMB) and the large-scale structure \cite{Bennett_2003a, Tegmark_2004a}. The theoretical framework with the consideration of DE has become a shortcoming of General Relativity (GR). So, there is a significant interest in exploring an alternative gravity to GR. In this framework, it is usually necessary to include additional terms in the Einstein-Hilbert action, which preserves the GR predictions at local scales. It introduces corrections at cosmological scales, which resulted in the expanding and accelerating behavior of the Universe at a late time \cite{Nojiri_2011a, Capozziello:2011et}. Theoretically, explaining the accelerated expansion of the Universe can be accomplished in two approaches: first, by introducing the matter field in the dark energy sector, such as the canonical scalar field, and phantom field, or combining both into one unified model that leads to more complex models \cite{COPELAND_2006, Cai_2010}. Second, the gravitational sectors themselves can be modified \cite{Capozziello:2011et, Nojiri_2011a, De_Felice_2010, Lobo_2009}.\\
 
We shall discuss some of the dark energy cosmological models in the modified theories of gravity such as $f(R)$ gravity \cite{Nojiri2006, NOJIRI2007238a}, $f(T)$ gravity \cite{Linder:2010py, Baojiu2011a}, $f(R, \mathcal{T})$ gravity \cite{Harko2011a} and $f(\mathcal{G})$ gravity \cite{Nojiri2005a}, where $R$, $T$, $\mathcal{T}$ and $\mathcal{G}$ represents the Ricci scalar curvature, torsion scalar, trace of matter energy-momentum tensor and Gauss-Bonnet term respectively. The cosmological solutions of these models explained the accelerated expansion of the Universe. The focus in this work is based on the torsion-based gravitational theory. The torsion formalism is equivalent to GR, referred to as the teleparallel equivalent of general relativity (TEGR), up to a boundary term difference \cite{Maluf:1994j, Arcos:2004tzt, Einstein28}. Accordingly, the corresponding Lagrangian $T$ is calculated using contractions of the torsion tensor, just as in Einstein-Hilbert Lagrangian $R$ is calculated using contractions of the curvature (Riemann tensor). Extending $T$ to a Lagrangian function, one can construct the modified gravity $f(T)$ starting from TEGR \cite{Ferraro2007a}. Even though TEGR is identical to GR in equations, $f(T)$ gravity does not follow the same rules as in $f(R)$ gravity. There are many interesting and novel implications for cosmology in $f(T)$ gravity \cite{Linder:2010py, Chen:2010va, Dent_2011a}. Additionally, one can construct higher-order torsion gravity, such as the $f(T, T_\mathcal{G})$ paradigm \cite{Kofinas:2014owa, Kofinas:2014daa}, by using the TEGR paradigm. This paradigm also exhibits interesting cosmological characteristics \cite{Kofinas:2014aka, Lohakare_2023a}. Another modification of TEGR involves the introduction of Lagrange multiplier into a Weyl-Cartan geometry to extend it through the Weitzenb$\ddot{o}$ck condition \cite{Haghani_2012aa, Haghani_2013a}.\\

The matter-coupled modified gravity theory can also be considered in TEGR. 

In $f(T)$ gravity, it is shown that power-law solutions can be found for the function $f(T)$, including those during phantom phase leading to a big rip singularity \cite{Setare_1110.3962}. Analyzing singularity analysis of differential equations, Palanthasis et al. \cite{Paliathanasis_1606.00659} studied isotropic, homogeneous, and anisotropic universes with dust and radiation in $f(T)$ gravity. One among the first proposed theories is the $f (T, \mathcal{T})$ gravity [Harko et al. \cite{Harko_2014a}]. Considering the torsion scalar $T$ and the trace of the energy-momentum tensor $\mathcal{T}$, a part of the gravitational Lagrangian can be described arbitrarily. In contrast to the theories based on curvature or torsion, $f (T, \mathcal{T})$ appears to follow an entirely different formalism. Momeni and  Myrzakulov \cite{Momeni_2014} have performed the cosmological reconstruction in $f(T, \mathcal{T})$ gravity and explained the accelerated expansion of the Universe. Farrugia and Jackson  \cite{Jackson2016a} have investigated the growth factor for subhorizon modes during late times in $f(T, \mathcal{T})$ gravity. Junior et al. \cite{Junior_2016} have studied the reconstruction in the gravitational action of the $\Lambda$CDM model and have done a brief analysis of their stability. Pace and Jackson \cite{Pace_2017} have derived a working model for the Tolman-Oppenheimer-Volkoff equation for quark star systems within the modified $f(T, \mathcal{T})$ gravity class of models. During the inflation phase and the late-time dark energy-dominated phase, Nassur et al. \cite{Nassur_2015a} examined the properties of various $f (T, \mathcal{T})$ models. %{\color{blue} A covariant formulation of scalar-torsion gravity is considered in the action considered by Manuel Hohmann et al. \cite{PhysRevD.97.104011}; when a flat but nontrivial spin connection is included, the Lorentz invariant formulation is possible.}\\

The dynamical system approach is an effective tool to examine the entire asymptotic behavior of the cosmological model, and it allows us to avoid the challenge of solving non-linear cosmological equations. This technique also describes the overall dynamics of the Universe by analyzing the local asymptotic behavior of critical points of the system and connecting them to the major cosmological epochs of the Universe. For example, the radiation and matter-dominated periods correlate to saddle points, but late-time (the dark energy sector) dominance normally corresponds to a stable point. An interesting feature of the cosmological model is its behavior with a focus on late-time stable solutions in a dynamic manner. In addition to the initial conditions and the evolution of the Universe, the phase-space and stability analysis reveals the global features of the cosmological scenario, which cannot be achieved from the initial conditions. Several modified gravity-based cosmological has been analyzed by using the dynamical system approach \cite{Dent_2011a, Mirza_2017, Bahamonde_2019aqa, Franco2021aca, Granda_2020a, Gonzalez-Espinoza:2020jss, Kadam:2022lgq, Narawade_2022a, Duchaniya_2023, Lohakare_2023_40, Agrawal_2023}. Motivated by the recent work of Duchaniya et al. \cite{Duchaniya_2022} on dynamical system analysis, in this paper, we investigate the stability of $f(T,\mathcal{T})$ through a dynamical system analysis approach. We explore the dynamics of the critical points for the better determination of its viability, which could correlate with the real behavior of the Universe. The graphical behavior further emphasizes the findings of the study. The arrangement of the manuscript is as follows: In Sec. \ref{SEC-II}, we discuss the $f(T,\mathcal{T})$ gravitational modification in a cosmological framework. The dynamical systems approach is used in Sec. \ref{SEC-III} to analyze the stability of the $f(T,\mathcal{T})$ cosmological model. Finally, the conclusions are highlighted in Sec. \ref{SEC-IV}.

%%%%%%%%%%%%%%%%%%%%%%%%%%%%%%%%%%%%% 
\section{Mathematical formalism} \label{SEC-II}
In this section, we shall discuss the development of cosmological equations in the context of $f(T,\mathcal{T})$ gravity. In the GR framework, the metric tensor $g_{\mu \nu}$ has been used, whereas the tetrad fields $e^{A}_{\mu}$ are used as a dynamical variable in the teleparallel framework. To note, using Greek notation, the space-time coordinates are indexed, whereas using capital Latin notation, the tangent space-time coordinates are indexed. We can write the metric tensor as,
\begin{equation}\label{1}
g_{\mu \nu}=\eta_{AB} e_{\mu}^{A} e_{\nu}^{B} \,,    
\end{equation}
where $\eta_{AB}$ represents the Minkowski space-time and the tetrad field satisfies the orthogonality conditions $e^{\mu}_{\,\,\,\, A}  e^{B}_{\,\,\,\,\mu}=\delta_A^B$. In order to describe $f(T,\mathcal{T})$ gravity, the Weitzenb$\ddot{o}$ck connection has been used as,  
\begin{equation}\label{2}
\hat\Gamma^{\lambda}_{\nu \mu}\equiv e^{\lambda}_{A} (\partial_{\mu} e^{A}_{\nu}+ \omega^{A}_{\,\,\, B \mu} e^{B}_\nu).
\end{equation}
where $\omega^{A}_{\,\,\, B \mu}$ is a flat spin connection in the theory incorporates Lorentz transformation invariance, which arises explicitly from the tangent space indices, contrasted with GR, whose spin connections are not flat \cite{misner1973gravitation} due to their tetrads. A tetrad-spin connection pair represents gravitational and local degrees of freedom in TG, and both contribute to the equations of motion of a system. Further, the torsion tensor, which is an anti-symmetric part of the Weitzenb$\ddot{o}$ck connection, can be expressed as, 
\begin{equation}\label{3}
T^{\lambda}_{\mu \nu}\equiv\hat\Gamma^{\lambda}_{\nu \mu}-\hat\Gamma^{\lambda}_{\mu \nu}=e^{\lambda}_{A} \partial_{\mu} e^{A}_{\nu}-e^{\lambda}_{A} \partial_{\nu} e^{A}_{\mu}.    
\end{equation}

Both diffeomorphisms and Lorentz transformations covariate with the torsion tensor. Subsequently assuming appropriate contractions of the torsion tensor, the torsion scalar can be expressed as, 
\begin{equation}\label{4}
T \equiv \frac{1}{4} T^{\rho \mu \nu} T_{\rho \mu \nu}+\frac{1}{2} T^{\rho \mu \nu} T_{\nu \mu \rho}-T_{\rho \mu}^{~~\rho} T^{\nu \mu}_{~~\nu}.
\end{equation}

The teleparallel Lagrangian $T$ constructs the action in teleparallel gravity. The notion of $f(T)$ gravity is to generalize $T$ to any arbitrary function $f(T)$, which is conceptually comparable to the generalization of the Ricci scalar $R$ in the Einstein-Hilbert action to a function $f(R)$. So, the action of $f(T,\mathcal{T})$ gravity can be given as, 
\begin{equation}\label{5}
S = \frac{1}{16 \pi G}\int d^{4}xe[T+f(T,\mathcal{T})+\mathcal{L}_{m}],    
\end{equation}
where $\mathcal{L}_{m}$ be the matter Lagrangian and $f(T,\mathcal{T})$ is an arbitrary function of torsion scalar $T$ and trace of energy momentum tensor $\mathcal{T}$. The gravitational constant is denoted as $G$. The tetrad determinant can be represented as $e=\text{det}[e^{A}_{\,\,\,\,\mu}]=\sqrt{-g}$ and by varying the action \eqref{5} with respect to the tetrad field, the gravitational field equation of $f(T,\mathcal{T})$ gravity can be given as, 
%\begin{widetext}
\begin{eqnarray}\label{6}
&&[e^{-1}\partial_{\mu}(e e^{\rho}_{A}S_{\rho}^{~\mu \nu})-e^{\lambda}_{A}T^{\rho}_{~\mu \lambda}S_{\rho}^{~ \nu\mu}](1+f_{T}) \nonumber \\ &&+e^{\rho}_{A}S_{\rho}^{~\mu \nu}[\partial_{\mu}(T)f_{TT} +\partial_{\mu}(\mathcal{T})f_{T\mathcal{T}}]+\frac{1}{4}e^{\nu}_{A}[T+f(T)]\nonumber \\ &&-f_\mathcal{T}\left(\frac{e^{\rho}_{A}T_{~\rho}^{~~\nu}+p e^{\rho}_{A}}{2}\right)=4 \pi G e^{\rho}_{A}T_{~\rho}^{~~\nu}.
\end{eqnarray}
%\end{widetext}

Here, we follow the notation, $f_{T}=\frac{\partial f}{\partial T}$, $f_{TT}=\frac{\partial^2 f}{\partial T^2}$, $f_{\mathcal{T}}=\frac{\partial f}{\partial \mathcal{T}}$, $f_{T\mathcal{T}}=\frac{\partial^2 f}{\partial T \partial \mathcal{T}}$. In addition, the total energy-momentum tensor can be denoted as $T_{~\rho}^{~~\nu}$  and the superpotential as $S_{\rho}^{~~\mu \nu}\equiv\frac{1}{2}(K^{\mu \nu}_{~~~\rho}+\delta^{\mu}_{\rho}T^{\alpha \nu}_{~~~\alpha}-\delta^{\nu}_{\rho}T^{\alpha \mu}_{~~~\alpha})$. The contortion tensor in the superpotential can be expressed as $K^{\mu \nu}_{~~~\rho}\equiv \frac{1}{2}(T^{\nu \mu}_{~~~\rho}+T_{\rho}^{~~\mu \nu}-T^{\mu \nu}_{~~~\rho})$. For the cosmological study, we consider the homogeneous and isotropic flat Friedmann-Lema\^{i}tre-Robertson-Walker (FLRW) space-time as, 
\begin{equation}\label{7}
ds^{2}=dt^{2}-a^{2}(t)[dx^2+dy^2+dz^2]\,,
\end{equation}
where $a(t)$ is the scale factor that represents the expansion rate in the spatial directions.. Using Eq. \eqref{4}, the torsion scalar can be obtained as
\begin{equation}\label{8}
T=-6H^{2} = -6 \left(\frac{\dot{a}(t)}{a(t)}\right)^2,   
\end{equation}
where $a(t)$ is the scale factor, and the corresponding tetrad field can be written as $ e^{A}_{\mu}\equiv diag(1, a(t), a(t), a(t))$. Now, we can obtain the field equations of $f(T,\mathcal{T})$ gravity [Eq. \eqref{6}] as,

\begin{eqnarray}
3H^2&=&8\pi G \rho_m-\frac{1}{2}(f+12 H^{2}f_T)+f_{\mathcal{T}}(\rho_m+p_m), \label{9}\\
\dot{H}&=&-4\pi G(\rho_{m}+p_{m})-\dot{H}(f_{T}-12 H^{2}f_{TT}) \nonumber\\ &&-H(\dot{\rho}_{m}-3\dot{p}_{m}) f_{T \mathcal{T}}-f_{\mathcal{T}}\left(\frac{\rho_m+p_m}{2}\right).\label{10} 
\end{eqnarray}
An over dot on the Hubble parameter $H$ denotes the ordinary derivative with respect to time $t$ and the trace of the energy-momentum tensor, $\mathcal{T}=\rho_{m}-3 p_{m}$. Here, $p_{m}$ represents the pressure of matter, whereas the equivalent energy density terms are represented by $\rho_{m}$. It is noteworthy to mention here that the matter sectors make up the overall energy-momentum tensor. Next, the modified Friedmann Eqs. \eqref{9}-\eqref{10} are given as,
\begin{eqnarray}
3 H^{2}&=& 8 \pi G(\rho_{m}+\rho_{DE}), \label{11}\\
-\dot{H}&=&4 \pi G(\rho_{m}+p_{m}+\rho_{DE}+p_{DE}). \label{12}
\end{eqnarray}

From Eqs. \eqref{9}- \eqref{12}, the expression of energy density ($\rho_{DE}$) and pressure ($p_{DE}$) for the dark energy sector can be obtained,  
\begin{widetext}
\begin{eqnarray}
\rho_{DE}&\equiv&-\frac{1}{16 \pi G}\left[f+12 H^{2}f_{T}-2 f_{\mathcal{T}}(\rho_m+p_m)\right], \label{13} \\
p_{DE}&\equiv& (\rho_m+p_m)\left[\frac{1+\frac{f_{\mathcal{T}}}{8 \pi G}}{1+f_{T}-12 H^{2}f_{TT}+H \dfrac{d \rho_{m}}{dH}(1-3 c^{2}_{s})f_{T \mathcal{T}}}-1\right]+ \frac{1}{16 \pi G}[f+12 H^{2}f_{T}-2 f_{\mathcal{T}}(\rho_m+p_m)].\label{14} 
\end{eqnarray}
\end{widetext}

Further, the effective dark energy equation of state (EoS) parameter can be derived,
\begin{equation} \label{15} 
\omega_{DE}= \frac{p_{DE}}{\rho_{DE}}.    
\end{equation}

The fulfilment of the fluid equations
\begin{eqnarray}
\dot{\rho}_m+3 H \rho_{m}=0,\label{16} \\
\dot{\rho}_{DE}+3 H (\rho_{DE}+p_{DE})=0. \label{18} 
\end{eqnarray}
 and the total EoS parameter
\begin{equation} \label{19} 
\omega_{tot}= \frac{p_{m}+p_{DE}}{\rho_{m}+\rho_{DE}}\equiv -1-\frac{2\dot{H}}{3 H^{2}},    
\end{equation}
It is well known that there is a direct relationship between the deceleration parameters $(q)$ and total EoS parameter $(\omega_{tot})$ as follows,
\begin{equation}\label{20} 
q = \frac{1}{2}(1+3 \omega_{tot})
\end{equation}
To note $q<0$ indicates the accelerating behavior of the Universe, whereas for the decelerating behavior $q>0$. Now, the density parameters are obtained as,
\begin{equation}\label{21}
\Omega_{m}=\frac{8 \pi G \rho_{m} }{3 H^{2}}, \hspace{0.8cm} \Omega_{DE}=\frac{8 \pi G \rho_{DE} }{3 H^{2}}  
\end{equation}
satisfying
\begin{equation}\label{22}
\Omega_{m}+\Omega_{DE}=1.    
\end{equation}

\section{The Dynamical system framework} \label{SEC-III}
The dynamical system provides certain guidelines for the evolution of the system and the potential future behavior of cosmological models. The dynamical system may be depicted as an equation of the type  $X^{'}=f(X)$, where $X$ is the column vector and $f(X)$ is the equivalent column vector of the autonomous equations. The prime represents the derivative with respect to $N= |\text{ln}\,\,a(t)|$. The general form of the dynamical system for the modified FLRW equations is defined by Eqs. (\ref{9}-\ref{10}) can be obtained in this approach. We introduce the dimensionless variables,
\begin{eqnarray}\label{23}
x&=&-\frac{f}{6 H^{2}}, \hspace{0.8cm} y=-2f_{T}, \hspace{0.8cm} u = \frac{\rho_{m} f_{\mathcal{T} }}{3 H^{2}}.   
\end{eqnarray}
with these dimensionless variables, the first Friedmann Eq. \eqref{9} becomes
\begin{equation}\label{24}
\Omega_{m}+x+y+u=1.    
\end{equation}
The density parameter at various stages of the Universe's evolutionary history can be expressed in the dynamical system variable, 
\begin{eqnarray}
\Omega_{DE}&=& x+y+u ,\label{25} 
\end{eqnarray}
Also, Eq. \eqref{10} can be rewritten in terms of dynamical variables as
\begin{equation}\label{28}
\frac{\dot{H}}{H^{2}}= \frac{-3+3x+3y+6 \rho_{m} f_{T \mathcal{T}}}{(2-y+24 H^{2} f_{TT})}.   
\end{equation}
From Eq. \eqref{23}, we construct the set of autonomous differential equations 
\begin{eqnarray}
\frac{dx}{dN}&=&\frac{3 u}{2}-(y+2 x)\frac{\dot{H}}{H^{2}}, \label{29}\\
\frac{dy}{dN}&=& 24 \dot{H} f_{TT}+6 \rho_{m} f_{T \mathcal{T}}, \label{30}\\
\frac{du}{dN}&=& -\frac{\rho_{m}^{2}f_{\mathcal{T} \mathcal{T}} }{H^{2}}-4\rho_{m} f_{\mathcal{T} T}  \frac{\dot{H}}{H^{2}}-3u-2u\frac{\dot{H}}{H^{2}}, \label{31}
\end{eqnarray}

We derive the EoS parameter and deceleration parameter in terms of dimensionless variables as,
\begin{eqnarray}
\omega_{DE}&=\frac{-6 H^2+6x H^2  +6y H^2 +12 H^2 \rho_{m} f_{T\mathcal{T}}+6 H^2(1-\frac{y}{2}+2 T f_{TT})}{(1-\frac{y}{2}+2 T f_{TT})(-6x H^2-6y H^2-2 f_{\mathcal{T}}\rho_{m})}, \label{34}\nonumber\\
\end{eqnarray}
\begin{eqnarray}
\omega_{tot}&=& -1-\frac{-1+x+y+2 \rho_{m} f_{T \mathcal{T}}}{(1-\frac{y}{2}+12 H^{2} f_{TT})}, \label{35}\\
q&=& -1-\frac{-3+3x+3y+6 \rho_{m} f_{T \mathcal{T}}}{(2-y+24 H^{2} f_{TT})}.\label{36}
\end{eqnarray}

To solve the autonomous system of differential Eqs. (\ref{29})-(\ref{31}), $f_{TT}$, and $f_{T \mathcal{T}}$ to be expressed either as a dynamical variable or in the form of dimensionless variables. This may be achieved by choosing some specific form of $f(T,\mathcal{T})$. Here, we have considered two forms of $f(T,\mathcal{T})$ and discussed the dynamical system analysis of two models.\\

\subsection{Model-I} \label{SEC-A}
Considering the dynamical system with the above dimensionless variables, we need to determine whether the modified gravity works as a model for the Universe. We consider the form of $f(T,\mathcal{T})$ \cite{Harko_2014a} as, 
\begin{equation}\label{37}
f(T,\mathcal{T})=\alpha T^{n} \mathcal{T}+\beta,   
\end{equation}
so that

\begin{eqnarray}\label{39}
f_{T}&=& \alpha n T^{n-1} \mathcal{T} = -\frac{y}{2},\nonumber\\ f_{TT}&=& \alpha n(n-1) T^{n-2}\mathcal{T} = -\frac{y(n-1)}{2T}, \nonumber\\ f_{T \mathcal{T}}&=&\alpha n T^{n-1}.
\end{eqnarray}
where the model parameters $\alpha$, $\beta$, and $n$ are constant \cite{Harko_2014a}. For this choice of $f(T,\mathcal{T})$, we have obtained the relation between dynamical variables $u=\frac{y}{2}$. The dynamical system is reduced to only two dynamical variables, $x$ and $y$. Subsequently, the autonomous system (\ref{29})-(\ref{31}) can be written respectively as,

\begin{eqnarray}
\frac{dx}{dN}&=&\frac{(2 x+y) (3 (n-2) y-6 x+6)}{(2-4 n) y+4}+\frac{3 y}{4}, \label{39}\\
\frac{dy}{dN}&=&-y \left(\frac{3 (n-1) ((n-2) y-2 x+2)}{(2 n-1) y-2}-3\right). \label{40}
\end{eqnarray}

We can express the dark energy EoS parameter, total EoS parameter and the deceleration parameter with respect to the dynamical variables as,  
\begin{eqnarray}
\omega_{DE}&=&\frac{4 x-6 (n-1) y}{((2 n-1) y-2) (2 x+3 y)} , \label{44}\\
\omega_{tot}&=& -\frac{(2 x + 3 y - 3 n y)}{(2 + y - 2 n y)}, \label{45}\\
q&=& \frac{7 n y-6 x-8 y+2}{-4 n y+2 y+4}.\label{46}
\end{eqnarray}

We solve the combined equations as described in Eqs. (\ref{39})-(\ref{40}) to obtain the critical points to analyze the dynamical features of the autonomous system. Subsequently, we obtain the stability condition and the cosmology to describe the evolutionary phase of the Universe. The existence of critical points and their cosmological parameters are given in Table \ref{TABLE-I}. For the system Eqs. (\ref{39})-(\ref{40}), we have obtained three critical points, and in the following section, we will go through the characteristics of each critical point and its possible relation with the evolutionary phase of the Universe.
\begin{widetext}
\begin{center}
\begin{table}[H]
    \caption{ The critical points and background parameters of the dynamical system. } % title of Table
    \centering % used for centering table
    \begin{tabular}{|c|c|c|c|c|c|c|c|c|} % centered columns (5 columns)
    \hline %inserts double horizontal lines
    C.P. & $x_{c}$ & $y_{c}$ & $q$ & $\omega_{tot}$ & $\omega_{DE}$ & $\Omega_{DE}$ & $\Omega_{m}$ & Exists for \\ [0.5ex] % inserts table %heading
    \hline\hline % inserts single horizontal line
    $A_{1}$  & $0$ & $0$ & $\frac{1}{2}$ & $0$ & undefined & $0$ & $1$ & Always\\
    \hline
    $A_{2}$ & $1$ &$0$ & $-1$ & $-1$ & $-1$ &  $1$ & $0$& Always\\
     \hline
    $A_{3}$ & $\frac{3 n-n^2}{n^2-6 n+3}$ & $ \frac{2}{3}\left(n- \frac{9n-3n^{2}}{3-6n+n^{2}}+ \frac{3n^{2}-n^{3}}{3-6n+n^{2}}\right)$ & $\frac{1+2n}{2-2n}$ & $\frac{n}{1-n}$ & $\frac{3+n(6-n)}{(n-1)(n+3)}$ & $\frac{-n^2-3 n}{n^2-6 n+3}$ & $\frac{2 n^2-3n+3}{n^2-6 n+3}$& \begin{tabular}{@{}c@{}} $n^2-6 n+3\neq 0$, \\ $5 n^2-8 n+3\neq 0$\end{tabular} \\
    [1ex] % [1ex] adds vertical space
    \hline %inserts single line
    \end{tabular}
     % is used to refer this table in the text
    \label{TABLE-I}
\end{table}
\end{center}
\end{widetext}

 \begin{itemize}
\item \textbf{Critical Point $A_{1}$}: The critical point $A_{1}$ exists for all values of the free parameters and the density parameter, $\Omega_{m}=1$. The value of the total EoS parameter vanishes and therefore, it is the same as that of the EoS parameter of the matter-dominated phase, and this critical point always exists. The dark energy EoS could not be determined for the critical point. Since the deceleration parameter is positive, there is no sign of cosmic acceleration. The respective eigenvalues of the Jacobian matrix for this critical point are defined below. The presence of both positive and negative sign eigenvalues means the critical point shows unstable node behavior if $n<0$. If $n>0$, then it shows unstable saddle behavior according to the linear stability theory.
 \begin{align*}
 \{3,-3n\}\,. \nonumber    
\end{align*}   
\end{itemize}
 %%%%%%%%%%%%%%%%%%%%%
\begin{itemize}
\item \textbf{Critical Points $A_{2}$:} The critical point $A_2$ appears in the dark energy era, and the density parameter provides $\Omega_{DE}=1$. The $\omega_{tot}=\omega_{DE}=-1$ and the deceleration parameter $q=-1$ indicates the late time cosmic acceleration of the Universe. Also, the Hubble rate is constant for these critical points, i.e., $\dot{H}=0$; therefore, the expansion keeps accelerating close to the critical point. The signature of the eigenvalues of this critical point is negative, which means this critical point shows stable node behavior for any choice of model parameters. 
\begin{align*}
 \{-3,-3\}\,. \nonumber    
\end{align*}
\end{itemize}
%%%%%%%
\begin{itemize}
\item \textbf{Critical Point $A_{3}$:} The dominance of different eras of the Universe for this critical point depends on the different range of model parameter $n$. We have presented background parameters value in Table \ref{TABLE-I}, which depend on the parameter $n$. The value of background parameters at $n=0$ is the same as critical point $A_{1}$. That means this critical also shows the matter-dominated phase of the Universe at $n=0$. According to the observations, the deceleration parameter shows the accelerated phase when $q<0$ and the decelerated phase when $q>0$ of the Universe. For this critical point, we have obtained the range of the model parameter $n$ to study different phases of the Universe. At $n<-\frac{1}{2}\lor n>1$, the deceleration parameter shows the accelerated phase of the Universe. For this range of $n$, the total EoS parameter satisfies the condition $\omega_{tot}<-\frac{1}{3}$. The total EoS parameter shows the phantom and quintessence regions of the Universe for $n>1$ and $n<-\frac{1}{2}$, respectively. The behavior of the eigenvalues depends on the model parameter $n$. This critical shows stable behavior for the range of model parameter $1<n\leq 2.14$. The signature of both eigenvalues is negative for this range. This critical point shows the overall dynamics of the Universe for different choices of model parameter $n$.
\begin{eqnarray*}
 \left\{\frac{3 \left(2 n^3-9 n^2-\sqrt{A}+3\right)}{2 (n-1) (5 n-3)},\frac{3 \left(2 n^3-9 n^2+\sqrt{A}+3\right)}{2 (n-1) (5 n-3)}\right\}   
\end{eqnarray*}
Where $A =4 n^6-36 n^5+101 n^4-120 n^3+78 n^2-36 n+9 $ 
\end{itemize} 

\begin{figure}[hbt!] 
    \centering
    \includegraphics[width=62mm]{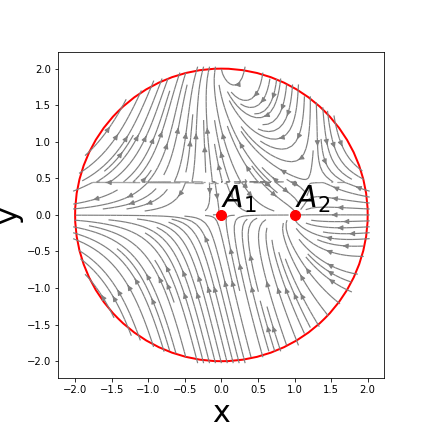}
    \includegraphics[width=62mm]{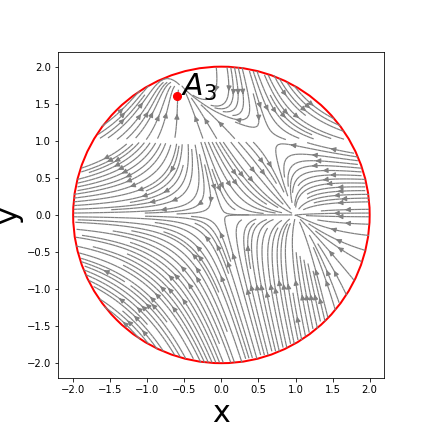}
    \includegraphics[width=62mm]{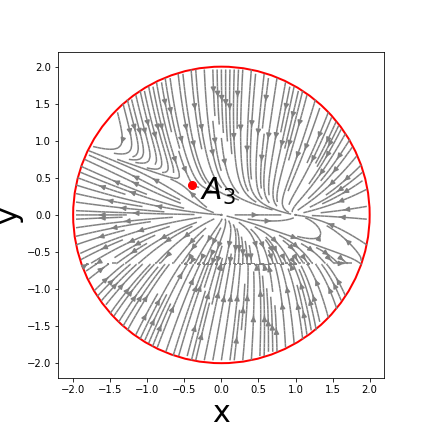}
    \caption{ $2$D Phase portrait for the dynamical system of {\bf Model-I}, (i) {\bf Upper panel:} Red dot denotes the critical point $A_{1}(0,0)$ and $A_{2}(1,0)$; (ii) {\bf Middle panel:} Red dot denotes the critical point $A_{3}(-0.6,1.6)$ for the model parameter $n=1.5$; (iii) {\bf Lower panel:} Red dot denotes the critical points $A_{3}(-\frac{2}{5}, \frac{2}{5})$ for the model parameter $n=-1$.} \label{Fig1}
\end{figure}

The phase portrait is an effective way to visualize the dynamics of the system as it shows the typical paths in the state space. The phase space of the critical points can be determined by setting the proper value for the parameters. Fig. \ref{Fig1} depicts the $2D$ phase space of the dynamical system specified in Eqs. (\ref{39})-(\ref{40}). The stability of the model can be described through the phase portrait. According to the phase space diagram (Fig. \ref{Fig1}), the trajectories of the critical point $A_{1}$ are moving away from the critical point, which indicates the saddle or unstable behavior. The trajectories show the attracting behavior towards the critical point, $A_{2}$, and hence show the stability. In addition, this critical point appear in the de-Sitter phase and may imply the current accelerated phase of the Universe. The trajectories for the critical point $A_{3}$ are moving towards the critical point $A_{3}$ for the stability range of model parameter $n<-\frac{1}{2}\lor n>1$, which is shown in Fig. \ref{Fig1} (middle-panel) for $n=1.5$ and outside the stability condition of the model parameter $n=-1$, the trajectories for the critical point $A_{3}$ are moving away to the critical point $A_{3}$, which means that outside the stability condition, the critical point shows unstable behavior, which is presented in Fig. \ref{Fig1} (lower-panel).

We have plotted the background dynamical parameter plot for the Universe's two different regions (Phantom and Quintessence). These regions are described from the total equation of the state parameter value for different choices of $n$. The Phantom region is defined for the range $n>1$, and the Quintessence region shows for the range $n<-\frac{1}{2}$. Here, we have drawn a plot for a specific choice of model parameter $n$; this particular choice of $n$ satisfies the condition of the Phantom and Quintessence regions.

\begin{figure}[H]
      \centering
      \includegraphics[width=70mm]{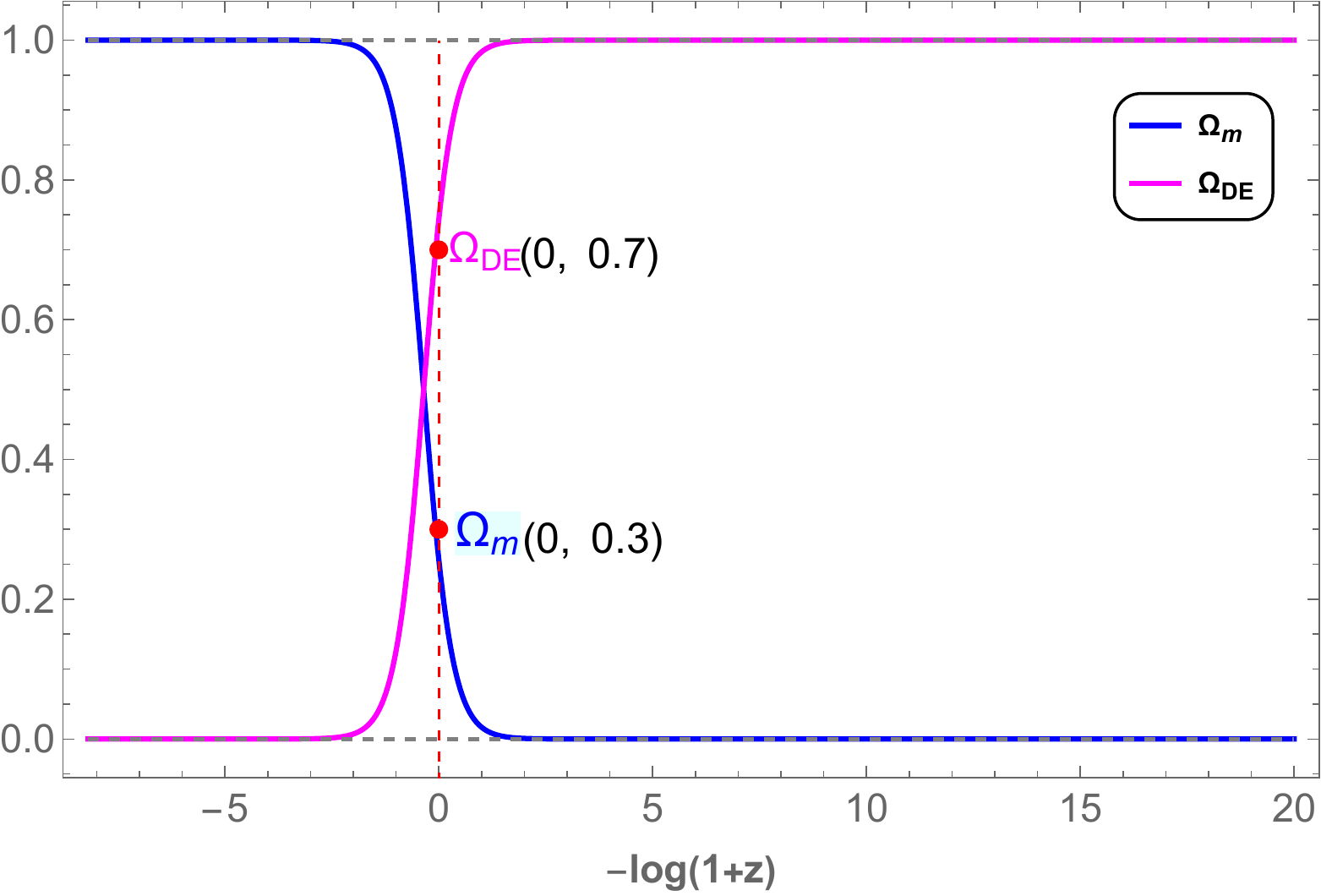}
      \includegraphics[width=70mm]{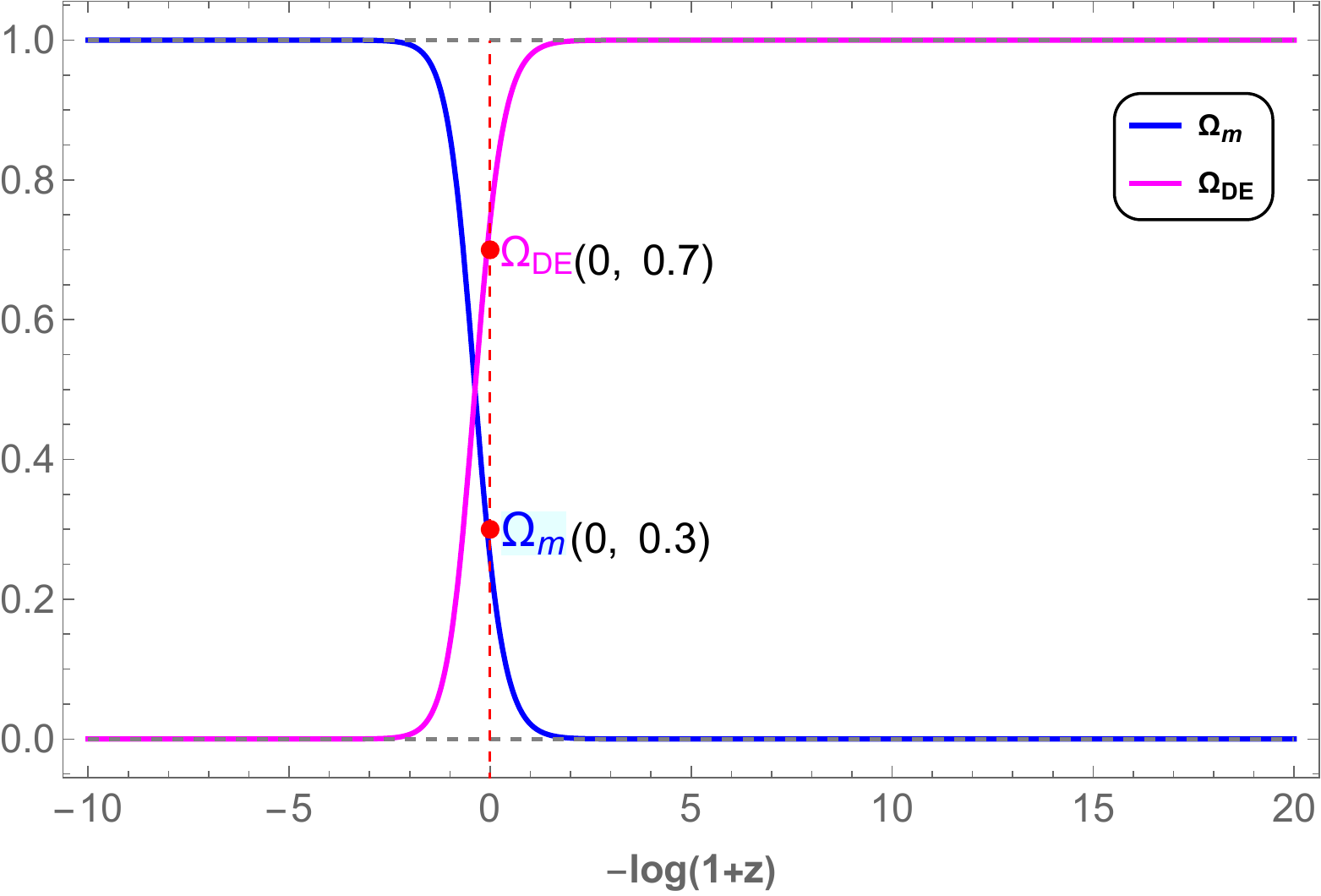}
    \caption{Evolution of density parameters for {\bf Model--I}. The initial conditions $x = 10^{-9}$, $y =10^{-8} $, and (\textbf{Upper panel n=1.5 Phantom region} ) and (\textbf{Lower panel n=-1 Quintessence region}). The vertical dashed red line denotes the present time.} \label{Fig2}
\end{figure}

  \begin{figure}[H]
      \centering
      \includegraphics[width=70mm]{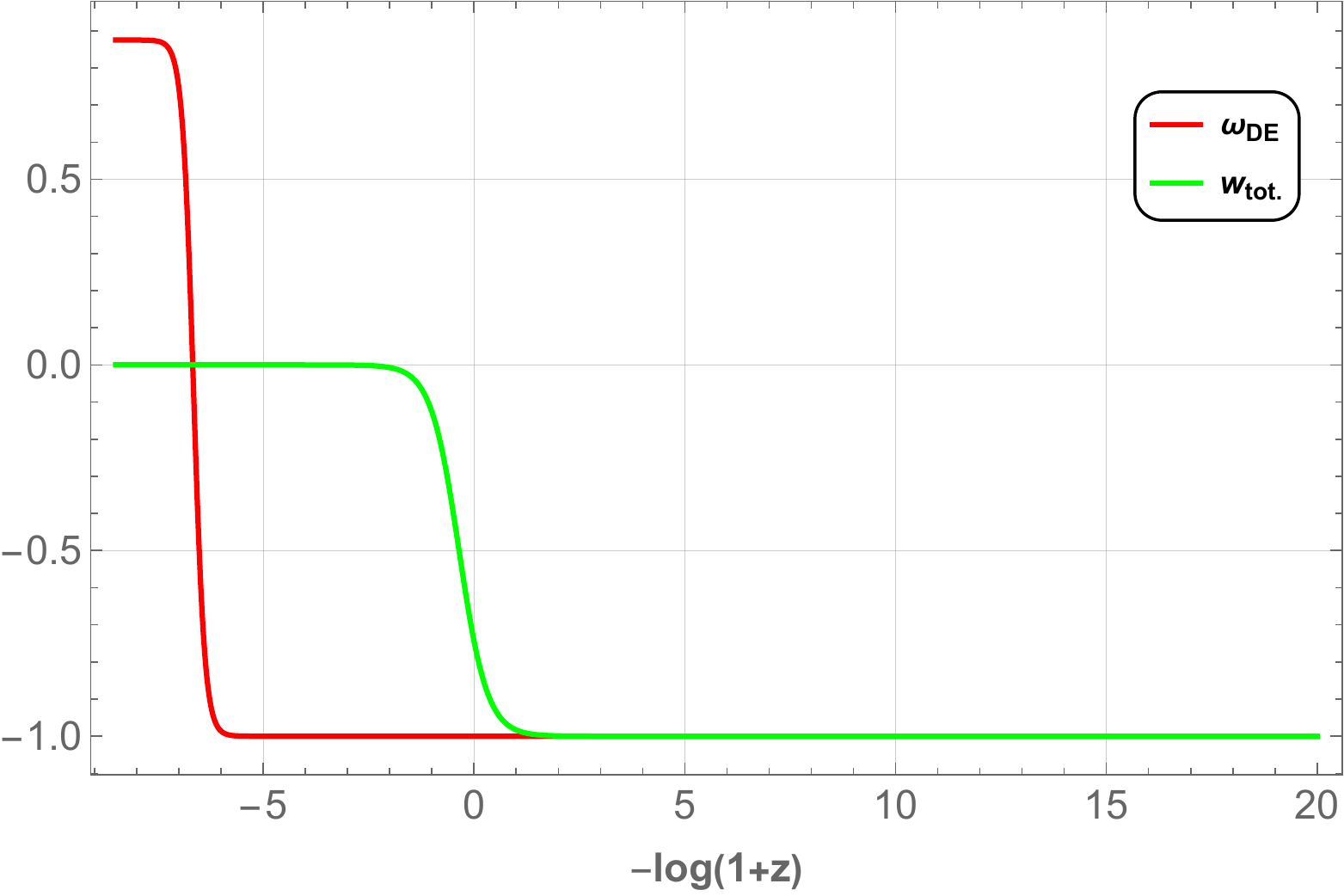}
      \includegraphics[width=70mm]{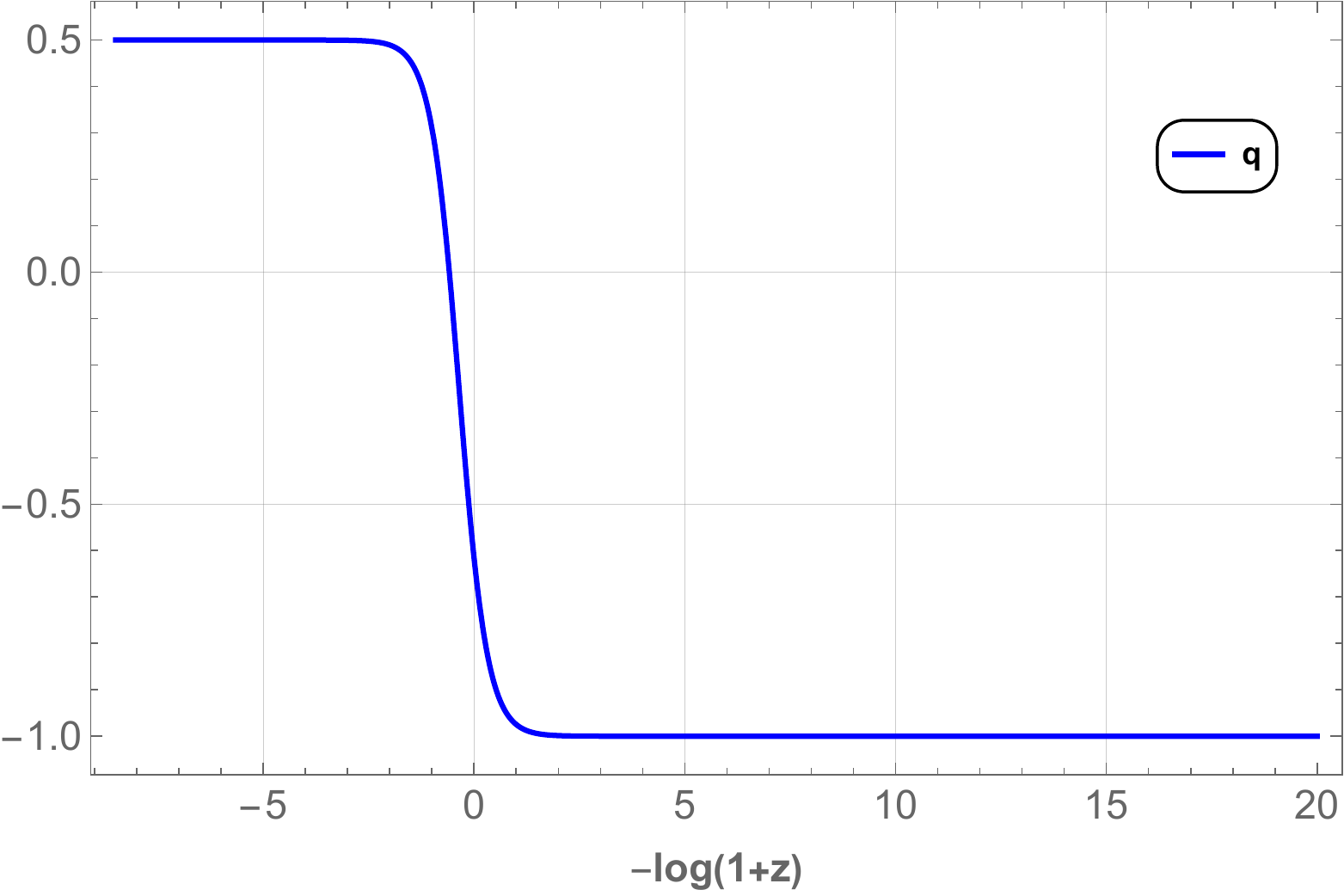}
      \caption{(\textbf{Upper panel}) Evolution of total EoS parameter (Green) and dark energy EoS parameter (Red); (\textbf{Lower panel}) Evolution of deceleration parameter (blue). The initial conditions are same as in Fig. \ref{Fig2} and $n=1.5$ (\textbf{Phantom region}) .} \label{Fig3}
  \end{figure}

    \begin{figure}[H]
      \centering
      \includegraphics[width=70mm]{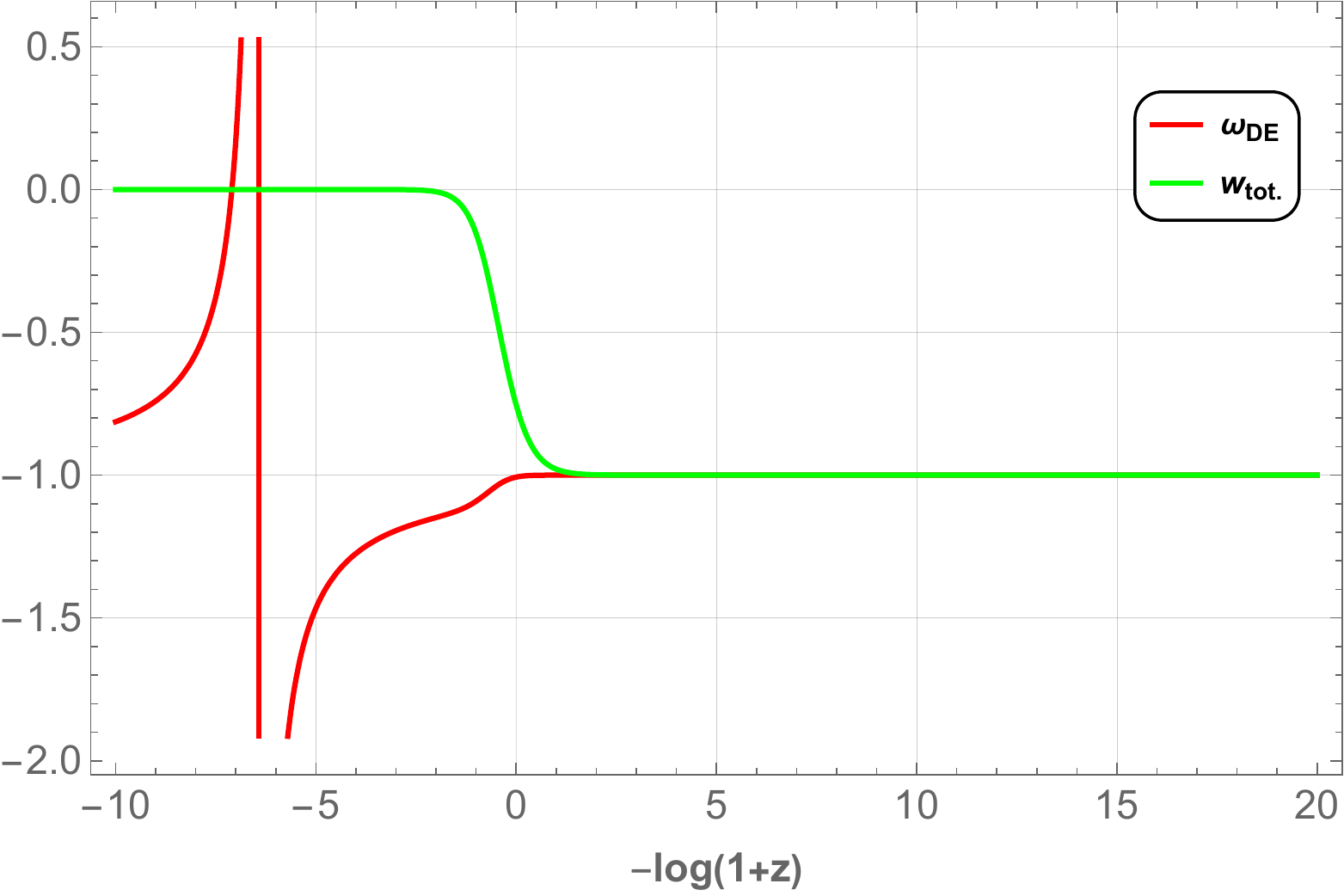}
      \includegraphics[width=70mm]{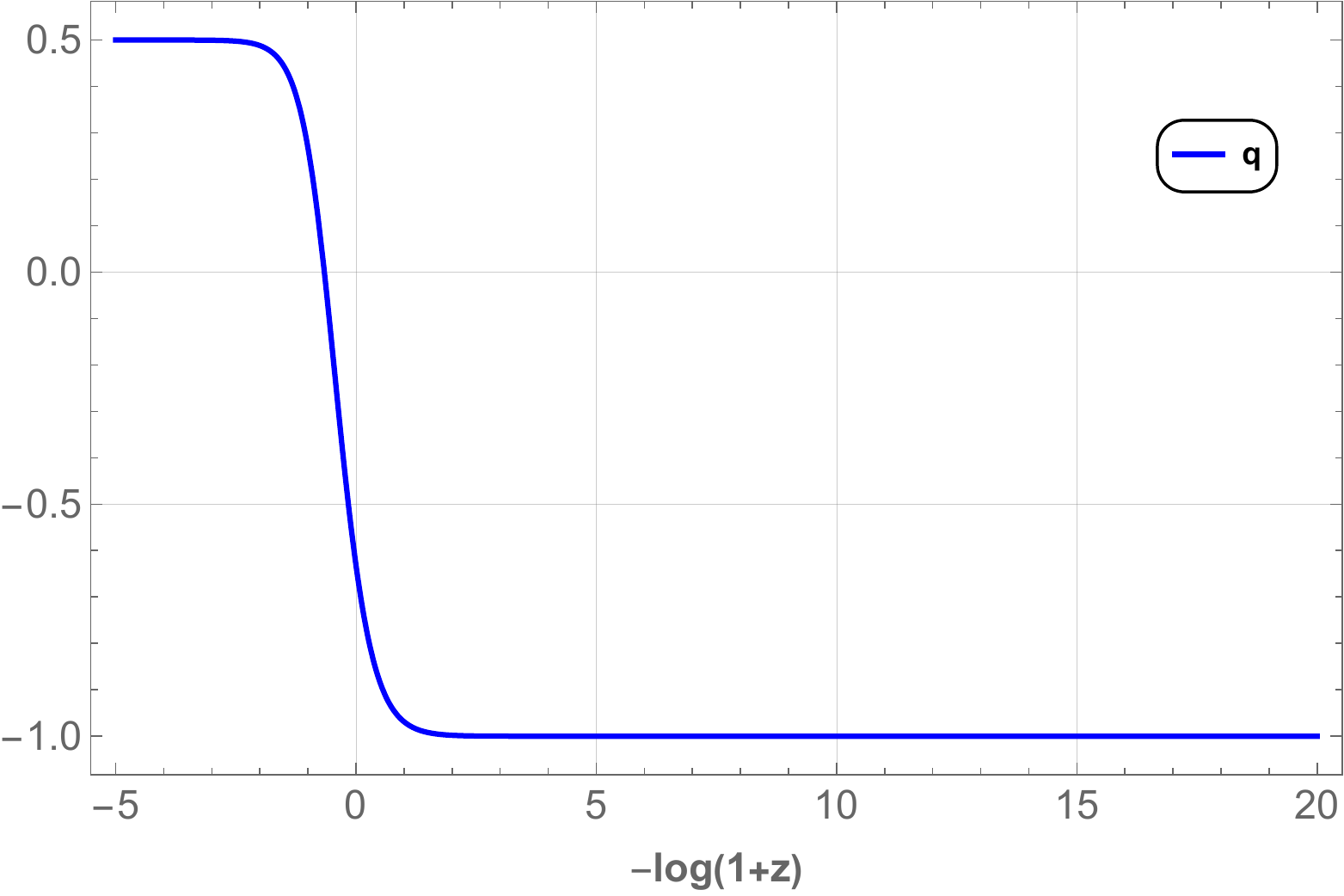}
      \caption{(\textbf{Upper panel}) Evolution of total EoS parameter (Green) and dark energy EoS parameter (Red); (\textbf{Lower panel}) Evolution of deceleration parameter (blue). The initial conditions are the same as in Fig. \ref{Fig2} and  $n=-1$ (\textbf{Quintessence region}).} \label{Fig4}
  \end{figure}

For both the Phantom and Quintessence regions, the evolutionary behavior of the density parameters (Fig. \ref{Fig2}) in redshift $\left(N=ln(\frac{1}{1+z})\right)$ illustrates the domination of matter in the early Universe, and later approaches the dark energy sector. At present (vertical red dashed line), as the observation revealed, dark matter and dark energy predominate. Dark energy makes up around $0.7$ of the total energy, with the remaining $0.3$ being dark matter. For both regions, We obtain $\Omega_{DE}\approx0.7$ and $\Omega_{m}\approx0.3$. We observe from Fig. \ref{Fig3} (upper panel) that the total EoS parameter begins with the Phantom region of the Universe, falls to $0$ during the period when matter predominated, and finally approaches $-1$ as the role of dark energy becomes more significant. We also notice the dark energy EoS parameter and, at present, $\omega_{DE}\approx -1$, which is compatible with the present Planck Collaboration result [$\omega_{DE}(z=0)= -1.028 \pm 0.032$ \cite{Aghanim:2018eyx}].  In Fig. \ref{Fig3} (lower panel), the deceleration parameter shows a transition from deceleration to acceleration with the transition at $z=0.62$, which is consistent with the observational constraint $z_{trans.}=0.7679^{+0.1831}_{-0.1829}$ \cite{PhysRevD.90.044016a}. The present value of the deceleration parameter can be obtained as $q(z=0) \approx -0.59$, consistent with the visualized cosmological observations \cite{PhysRevResearch.2.013028}. The total EoS parameter in Fig. \ref{Fig4} (upper panel) starts with a matter value of $0$ and moves towards $-1$ when dark energy plays a role. We also see the EoS parameter for dark energy, currently $\omega_{DE}\approx -1$. That agrees with the current Planck Collaboration conclusion. The deceleration parameter in Fig. \ref{Fig4} (lower panel) exhibits a transition from deceleration to acceleration at $z=0.58$, which follows the observational constraint. The calculated current value of the deceleration parameter is $q(z=0) \approx -0.63$, following the depicted cosmological data. \\

\subsection{Model-II} \label{SEC-B}
For further investigation of the dynamical system, we consider another form of $f(T,\mathcal{T})$ \cite{Harko_2014a} as
\begin{equation}\label{47}
f(T,\mathcal{T})= \gamma T^{2}+ \delta \mathcal{T},   
\end{equation}
where  $\gamma$ and $\delta$ are free parameters. The framework of the model in terms of dynamical variables is as follows:
\begin{eqnarray}\label{48}
f_{T}=2 \gamma T  \equiv -\frac{y}{2}, \hspace{0.5cm} f_{TT}&=&2\gamma, \hspace{0.5cm} f_{\mathcal{T}}=\delta,\nonumber\\ f_{\mathcal{T}\mathcal{T}}&=&0, \hspace{0.5cm} f_{T \mathcal{T}}=0.
\end{eqnarray}

We have discovered the relationship between the dynamical variables $y=-4(x+\frac{u}{2})$ for this choice of $f(T,\mathcal{T})$. Only the dynamic variables $x$ and $y$ remain in the simplified dynamic system. The autonomous system is therefore represented by the Eqs. (\ref{29})-(\ref{31})  as follows:
\begin{eqnarray}
\frac{dx}{dN}&=&\frac{3 (x+y-1) (4 (\delta +1) x+\delta  (-y)+2 (\delta +y))}{2 (\delta +1) (3 y-2)}, \label{49}\\
\frac{dy}{dN}&=&-\frac{6 y (x+y-1)}{3 y-2}. \label{50}
\end{eqnarray}
In the form of a dimensionless variable, the EoS parameter and deceleration parameter may be represented as,
\begin{eqnarray}
\omega_{DE}&=&\frac{(\delta +1) (2 x-y)}{(3 y-2) (\delta +x+y)} , \label{54}\\
\omega_{tot}&=& \frac{2 x-y}{3 y-2}, \label{55}\\
q&=& \frac{1-3 x}{2-3 y}.\label{56}
\end{eqnarray}

In order to perform the dynamical analysis, the critical points of the system to be obtained from Eqs. (\ref{49})-(\ref{50}). The nature and stability of each point to be established by perturbing the system around these critical points using the eigenvalues of the underlying perturbation matrix. The two critical points obtained are given in Table \ref{TABLE-IV} along with its existence condition(s) and the corresponding value of the EoS parameters, deceleration parameter and density parameters.
\begin{widetext}
\begin{center}
\begin{table}[H]
    \caption{ The critical points and background parameters of the dynamical system. } % title of Table
    \centering % used for centering table
    \begin{tabular}{|c|c|c|c|c|c|c|c|c|} % centered columns (5 columns)
    \hline\hline %inserts double horizontal lines
    C.P. & $x_{c}$ & $y_{c}$ & $q$ & $\omega_{tot}$ & $\omega_{DE}$ & $\Omega_{DE}$ & $\Omega_{m}$ & Exists for \\ [0.5ex] % inserts table %heading
    \hline\hline % inserts single horizontal line
    $B_{1}$  & $x$ & $1-x$ & $-1$ & $-1$ & $-1$ & $\frac{1-3x}{2}$ & $\frac{1+3x}{2}$ & $-\delta +3 \delta  x+3 x-1\neq 0$\\
    \hline
    $B_{2}$ & $-\frac{\delta }{2 (\delta +1)}$ &$0$ & $\frac{5 \delta +2}{4 \delta +4}$ & $\frac{\delta }{2 \delta +2}$ & $\frac{\delta +1}{2 \delta +1}$ &  $\frac{\delta }{2 (\delta +1)}$ & $\frac{\delta +2 }{2 (\delta +1)}$& $\delta +1\neq 0$\\
    [1ex] % [1ex] adds vertical space
    \hline %inserts single line
    \end{tabular}
     % is used to refer this table in the text
    \label{TABLE-IV}
\end{table}
\end{center}
\end{widetext}

\begin{itemize}
 \item {\bf Critical Point $B_{1}$:} The total EoS and dark energy sector EoS parameter and the deceleration parameter assume the value $-1$, which indicates an accelerating de-Sitter phase of the Universe. The value of the dynamical parameters shows that against the critical points, it shows $\Lambda$CDM-like behavior. The solution of density parameters for the critical point $B_{1}$ is $\Omega_{DE}=\frac{1-3x}{2 }$, and  $\Omega_{m}=\frac{1+3x}{2}$. For $x=-\frac{1}{3}$, the density parameters reduced in the fully dominated dark energy sector $\Omega_{DE}=1$. The first eigenvalue of this critical point is zero, and others depend on the model parameter $\delta$ and dynamical variable $x$. We know that the linear stability theory fails to tell about stability when zero eigenvalue is presented along with positive eigenvalues. But in our case the second eigenvalue is negative when $\delta$, and $x$ satisfy the condition $x\in \mathbb{R}\land \left(\delta <-1\lor \delta >-\frac{2}{3}\right)$. For this condition, the eigenvalues of this critical point are negative real part and zero. The dimension of the set of eigenvalues for non-hyperbolic critical points equals the number of vanishing eigenvalues \cite{Coley:1999,aulbach1984}. As a result, the set of eigenvalues is normally hyperbolic, and the critical point associated with it is stable but cannot be a global attractor. In our case, the dimension of the set of eigenvalues is one, and only one eigenvalue vanishes. That means the dimension of a set of eigenvalues equals the number of vanishing eigenvalues. This critical point is consistent with recent observations and can explain the current acceleration of the Universe. The behavior of this critical point is a stable node.\\
 
 \begin{align*}
 \left\{0,\frac{3 (3 \delta -9 \delta  x-6 x+2)}{2 (\delta +1) (3 x-1)}\right\}. \nonumber    
\end{align*}
\end{itemize}
  
\begin{itemize}
\item {\bf Critical Point $B_{2}$:} The critical point $B_{2}$ is the origin of the phase space, which corresponds to a Universe where matter predominates, $\Omega_{m}=1$ for $\delta=0$. Since $\omega_{tot}=0$ = $\omega_{m}$ and the total EoS overlaps with the matter EoS, no acceleration occurs for physically permissible $\omega_{m}$ values. The phase space trajectory and the eigenvalues show the unstable behavior of the critical point for any choice of model parameter $\delta$. 
 \begin{align*}
 \left\{-\frac{3 (3 \delta +2)}{2 (\delta +1)},\frac{3 (3 \delta +2)}{2 (\delta +1)}\right\}. \nonumber    
\end{align*}
\end{itemize}

\begin{figure}[hbt!]
    \centering
    \includegraphics[width=59mm]{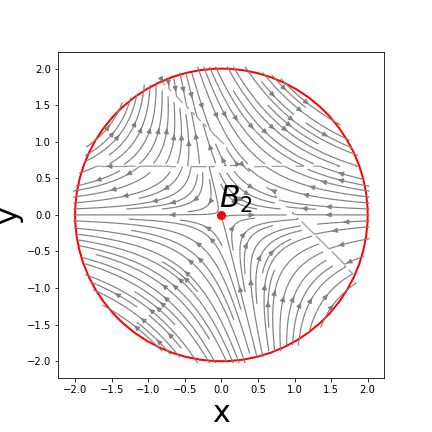}
    \includegraphics[width=59mm]{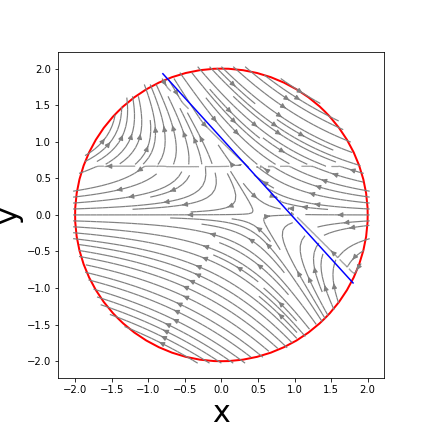}
     \caption{ $2D$ Phase portrait for the dynamical system of {\bf Model-II}, (i) {\bf Upper panel:} Red dot denotes the critical point  $B_{2}$;  (ii) {\bf Lower panel:} blue line denotes the critical points $B_{1}$.} \label{Fig5}
\end{figure}

 Further, to establish the result on the stability of the critical points, the $2D$ phase space trajectory of the dynamical system Eqs. (\ref{49})--(\ref{50}) are shown in Fig. \ref{Fig5}. The direction of the trajectory describes the stability behavior of the critical point. The trajectories of the phase space are moving away from the critical point $B_{2}$, which shows the saddle or unstable behavior. In comparison, the trajectory exhibits attractive behavior for the curve of critical points $B_{1}$. Moreover, the phase portrait shows that $B_{1}$ is a global attractor.

\begin{figure}[H]
     \centering
     \includegraphics[width=80mm]{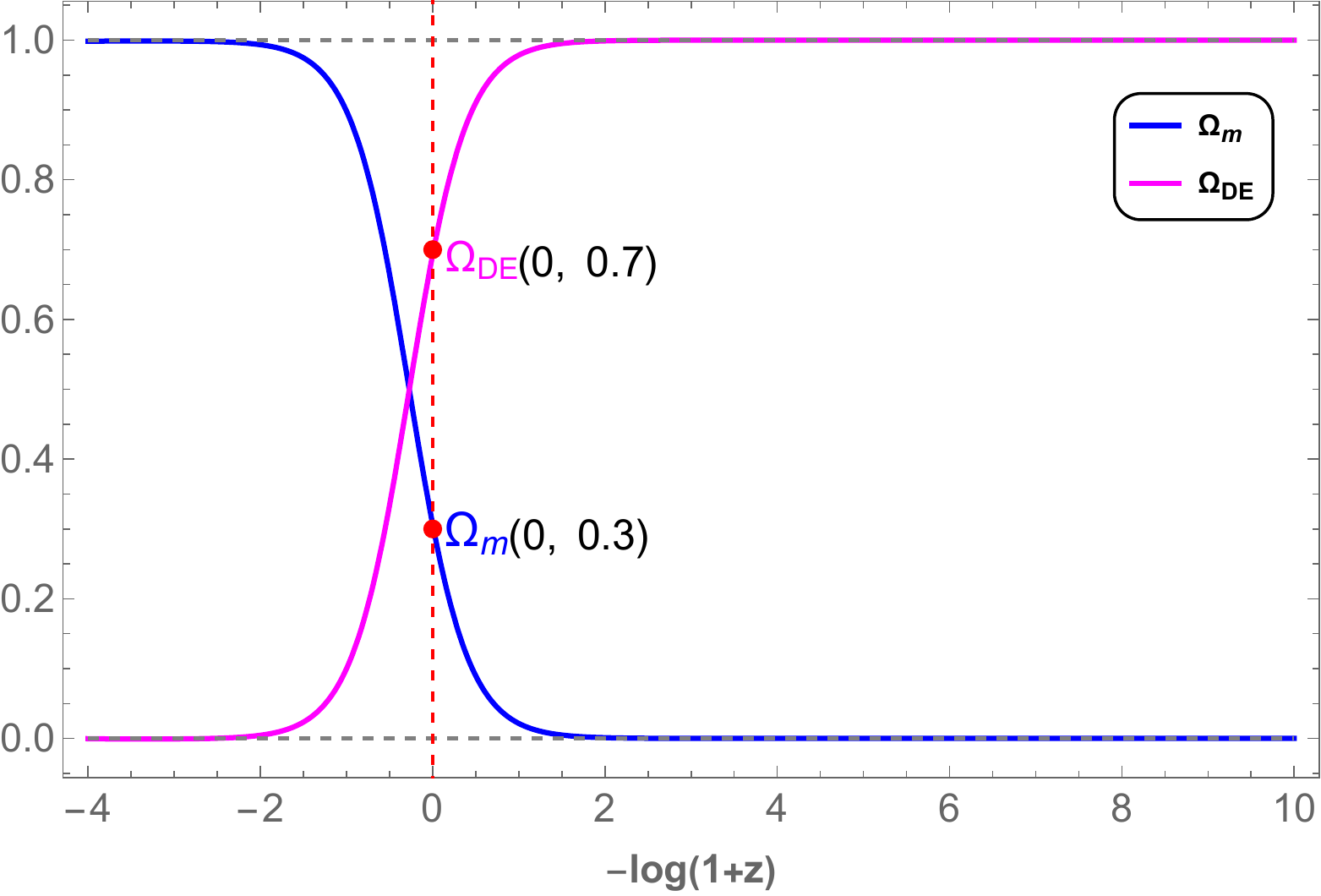}
     \caption{Evolution of density parameters for {\bf Model-II}. The initial conditions $x = 10^{-4}$, $y =10^{-5} $, $\delta=0.002$. The vertical dashed red line denotes the present time.}\label{Fig6}
\end{figure}

\begin{figure}[H]
     \centering
     \includegraphics[width=80mm]{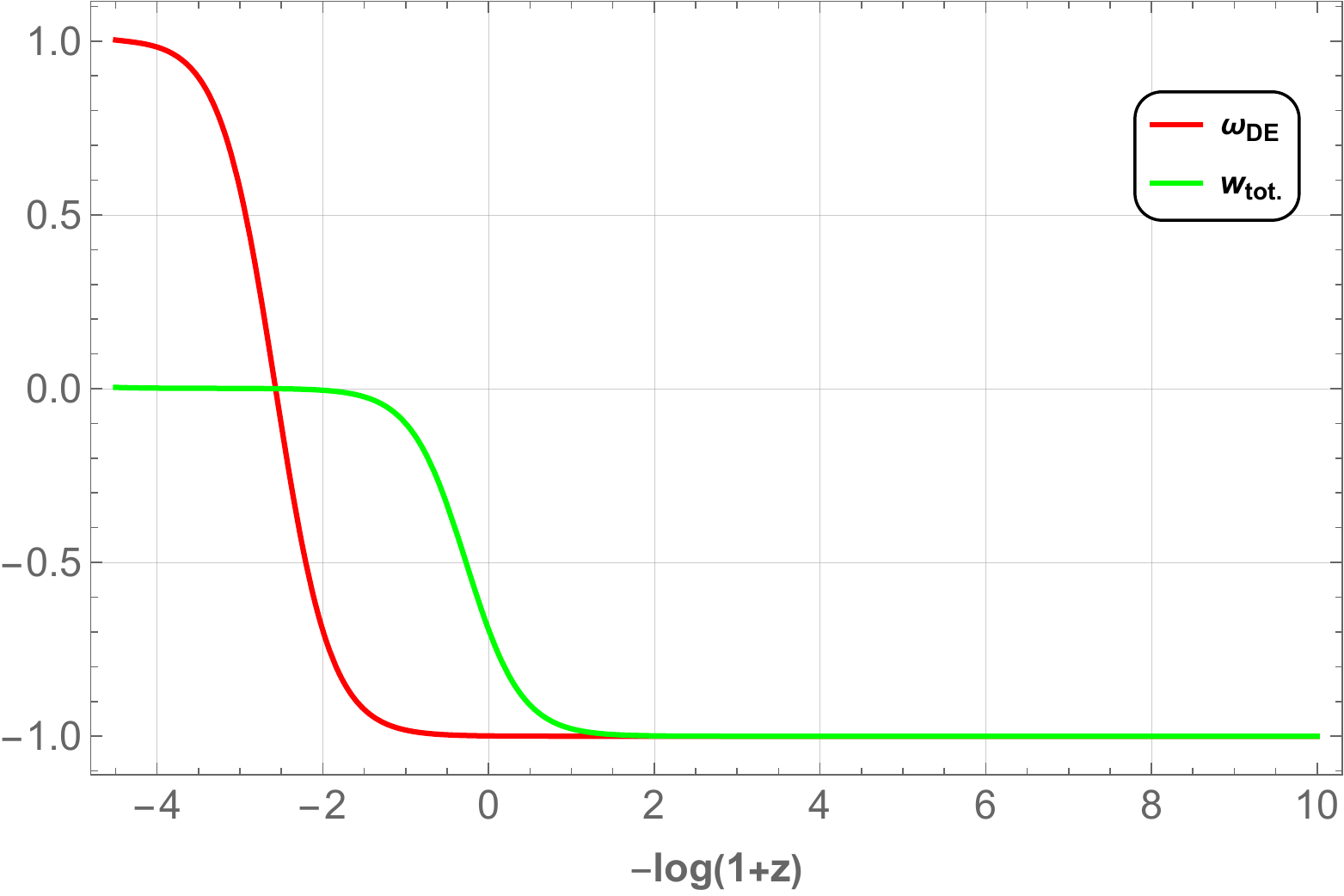}
     \includegraphics[width=80mm]{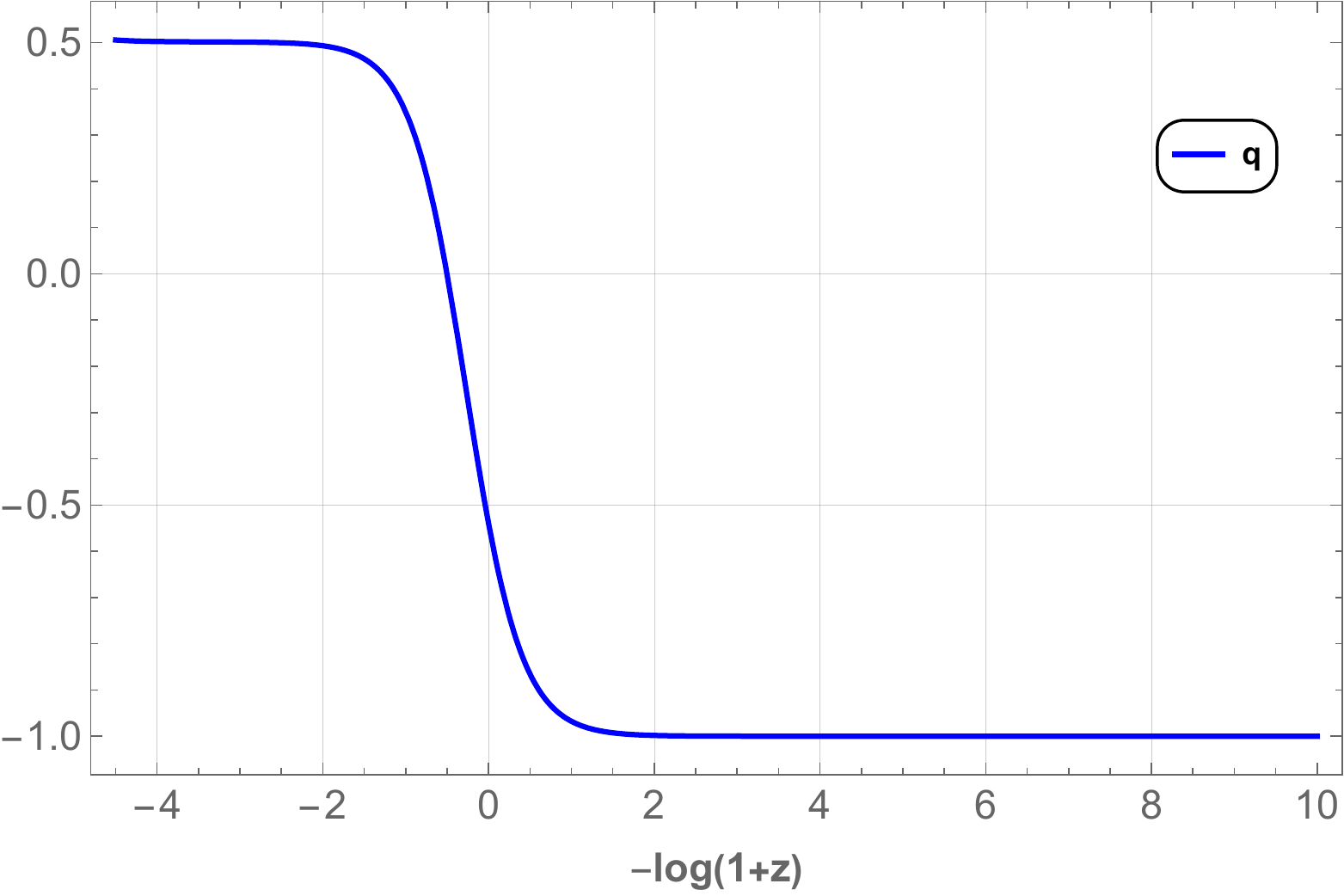}
      \caption{(\textbf{Upper panel}) Evolution of total EoS parameter (Green line) and dark energy EoS parameter (Red line); (\textbf{Lower panel}) Evolution of deceleration (Blue line). The initial conditions are same as in Fig. \ref{Fig6}.}  \label{Fig7}
\end{figure}

Fig. \ref{Fig6} shows how matter predominates early on and then gradually transitions into the de-sitter phase of the Universe. The present value for matter and dark energy density parameters are respectively $\Omega_{m}\approx0.3$ and $\Omega_{DE}\approx0.7$. The behavior of the EoS parameter has been given in Fig. \ref{Fig7}. The total EoS parameter starts with a matter value of  $0$ and finally approaches $-1$ as the role of dark energy becomes more substantial [Fig. \ref{Fig7} (upper panel)]. The dark energy EoS parameter approaches to $-1$, and the present value as suggested in Planck Collaboration is, $\omega_{DE}= -1.028 \pm 0.032 $ \cite{Aghanim:2018eyx}. The deceleration parameter shows transient behavior from deceleration to acceleration at the point $z=0.51$ with the present value noted as $q(z=0) \approx -0.56$ [Fig. \ref{Fig7}] (lower panel)] \cite{PhysRevResearch.2.013028}.

\section{Summary and Conclusion} \label{SEC-IV}
In conclusion, analysing the dynamical system with dimensionless variables in the context of modified teleparallel gravity with two forms of $f(T, \mathcal{T})$ has revealed several critical points and their cosmological implications.

In the first model $f(T,\mathcal{T})=\alpha T^n\mathcal{T}+\beta$, three critical points are obtained, out of which two critical points ($A_2, A_3$) are stable and the remaining critical point are unstable. The stable critical points are appeared in the de-Sitter phase, whereas the unstable behavior is noted in the matter-dominated phase of the Universe. The behavior of the critical points is obtained from the signature of the eigenvalues and further supported by the phase space portrait. It is clearly visible that the trajectories are going behavior of the unstable critical points, and the stable ones these are behaving as an attractor. The phase portrait and stability analysis provide insights into the dynamical features of the autonomous system. Trajectories illustrate the behavior of critical points, showing stable or unstable behavior based on the values of $n$. Additionally, the analysis reveals that the model parameters can lead to different cosmic scenarios, including matter domination, late-time acceleration, and the transition between these phases. The results align with observational constraints for the present values of $\omega_{DE}$ and $q$. For the critical points in the de-Sitter phase, both the values of the dark energy EoS parameter and deceleration parameter are $-1$, which confirms the accelerating model with the $\Lambda$CDM-like behavior.

In summary, the analysis of the dynamical system with an alternative form of $f(T,\mathcal{T})$ represented by $f(T,\mathcal{T})=\gamma T^2+\delta \mathcal{T}$ has provided important insights into the cosmological behavior of the Universe. The phase space trajectories illustrate the stability or instability of the critical points, with critical point $B_1$ being a global attractor and critical point $B_{2}$ showing unstable behavior. These results align with recent observations of the Universe's accelerating expansion. In this model also the stable critical points appear in the de-Sitter phase, and the value of the dark energy EoS parameter and deceleration parameter is $-1$. So the model appears to be $\Lambda$CDM like accelerating model. We have defined Phantom and Quintessence regions from the total equation of the state parameter value for different choices of $n$. The Phantom region is defined for the range $n>1$, and the Quintessence region shows for the range $n<-\frac{1}{2}$.

In this article, we have studied different phases of the Universe by analyzing the critical points. The behavior of critical points shows the evolutionary eras of the Universe, so we have obtained the exact solution of each critical point for both models. For critical point $A_{1}$, we have observed  the solution in terms of cosmic time $t$, i.e. $a(t)=c_{1} (\frac{3}{2}t)^\frac{2}{3}$. The solution to this critical point shows the matter-dominated phase of the Universe. For critical point $A_{2}$, we have obtained $a(t)=c_{2} e^{c_{1}t}$, which means $\dot{H}=0$, which is defined as the de-sitter phase of the Universe. For this point, we have $H=c_{1}$. The behavior of the solution shows the dark-energy-dominated phase of the Universe. The solution of the critical point $A_{3}$ is $a(t)=c_{0} t^{\frac{2(1-n)}{3}}$. The solution to this critical point depends on the model parameter $n$. The model parameter $n$ range is defined in branch \ref{SEC-A}. According to the model parameter choices $n$, the solution is a different Universe region. The solution shows that the Phantom region is defined for the range $n > 1$, and the Quintessence region is defined for the range $n < -\frac{1}{2}$. For the critical point $B_{1}$, we got $a(t)=c_{3} e^{c_{4}t}$. The solution shows a dark-energy-dominated phase of the Universe. Also, we have obtained the solution for critical point $B_{2}$, which is $c_{5} t^{\frac{4(\delta+1)}{3(3\delta+2})}$. This critical point shows the matter-dominated phase of the Universe for $\delta=0$. 

In both models, the total and dark energy EoS parameters merge, and the dominance of dark energy EoS is visible. The deceleration parameter shows the early deceleration and late time acceleration with the present value noted at $q_0=-0.59$ and $q_0=-0.56$ respectively for { \bf model I} and {\bf model II} and the transition is showing at $0.62$ and $0.51$. For both the models, the present value of the matter and dark energy density parameters are $\Omega_{m} \approx 0.3$ and $\Omega_{DE}\approx0.7$, and it fits the recent suggestions from cosmological observations. This analysis offers a comprehensive understanding of how modified gravity with the given form of $f(T, \mathcal{T})$ can describe the evolution of the Universe and its various phases based on the chosen model parameters.

\section*{Acknowledgements}
LKD acknowledges the financial support provided by the University Grants Commission (UGC) through Senior Research Fellowship UGC Ref. No.: 191620180688 to carry out the research work. SVL acknowledges the financial support provided by the University Grants Commission (UGC) through the Senior Research Fellowship  (UGC Ref. No.: 191620116597) to carry out the research work. BM acknowledges the support of IUCAA, Pune (India), through the visiting associateship program.

\bibliographystyle{utphys}
\bibliography{references}
\end{document}